\documentstyle[11pt,fullpage]{article}
\def \be{\begin{equation}}
\def \ee{\end{equation}}
\normalbaselines
\begin{document}
\begin{flushright}
TIFR/TH/94-13 PUB\\
October 1994 \\
(to appear in IJMP A)
\end{flushright}
\medskip
\begin{center}
{\Large \bf RECENT RESULTS IN NEUTRINO PHYSICS}
{\bf K. V. L. Sarma}\\[2mm] Tata Institute of Fundamental Research,
\\ Homi Bhabha Road, Bombay, 400 005, India\\ {\it (e-mail:
kvls@theory.tifr.res.in)}\\[10mm]
\end{center}
\begin{quotation}
\small{ This is a survey of the current experimental information on some
of the interesting issues in neutrino physics: neutrino species, neutrino
masses, neutrino magnetic moments, solar neutrinos, and the atmospheric
neutrino anomaly.}
\bigskip
\end{quotation}
\baselineskip=0.8cm

\centerline {\bf 1.~INTRODUCTION}
It is well-known that results in neutrino physics require patience and
enormous investments of experimental effort to obtain them. Hence not only
announcement of new results but even the confirmation of earlier results
or their revised versions become matters of considerable importance. In
the \break Non-Acclerator Particle Physics meeting held at Bangalore in January
1994 \cite{BAN}, many leading `neutrinoers' had presented a variety of
experimental observations. Some of these results have been superseded by
the ones reported at the High Energy Physics Conference at Glasgow in July
1994 \cite{GLA}. Here we draw together these results in the field of
neutrino physics, in the hope that such a compilation with a few
annotative remarks would be in the interests of information dissemination.
References to published literature are as of September 1994.

The topics are laid out under the following general heads: number of light
neutrino flavours, limits on the neutrino masses including those from
double beta-decay, and limits on neutrino magnetic moments obtained from
the data of laboratory-based experiments. Next we list the latest results
from the solar neutrino experiments, and discuss them in some detail. Of
crucial importance are the recent model-independent analyses which
determine the neutrino fluxes. The conundrum in terms of the near absence
of $^7$Be neutrino flux may be pointing towards nonstandard neutrino
physics, e.g., in terms of neutrino oscillations involving $\nu_{e}$ .
Towards the end, we look at the evidence for the deficit of the
muon-neutrino component in the atmospheric neutrinos. The zenith-angle
dependence observed recently by the Kamiokande group at the multi-GeV
neutrino energies, seems to favour neutrino oscillations involving $
\nu _{\mu }$ . For details regarding the specific experiments and other
standard definitions we refer the reader to other reviews.

\bigskip

\centerline {\bf 2.~NUMBER OF NEUTRINO SPECIES}

{}From the studies of the $Z$-lineshape at LEP, the number of species of the
standard light neutrinos is determined to be \cite{GLA}
\be N_{\nu }~=~ 2.988\pm 0.023~.                        \ee
Here the word `light' means anything which is $\ll (M_Z/2)~\approx ~$45
GeV. This result is quite justifiably regarded as a triumph for the
astro-particle physicists who arrived at the limit $N_{\nu }\leq 4$ on the
basis of the present mass fraction of $^4$He in the Universe \cite{STE}.

\bigskip

\centerline {\bf 3.~NEUTRINO MASSES}

All existing data are of course still consistent with massless neutrinos.
So a parade of negative results will follow:

{\bf (A)} ELECTRON NEUTRINO: Limit on the antineutrino mass comes from the
studies of the tritium beta spectrum close to the endpoint. The best upper
limit and the first result of the INR-KIAE collaboration \cite{BEL} was
given at the ICHEP-94 meeting \cite{BELL},
\be m(\bar {\nu }_e)~\leq ~ 4.5~{\rm eV~(95\%~CL)}~.\label{mass}     \ee
Although preliminary, it is an improvement over the earlier 7.2 eV limit
of the Mainz group \cite{MAI}. There is however cause for embarrassment.
The quantity that actually gets measured is the squared-mass of the
neutrino which comes out negative $m^2(\bar {\nu }_e)=-20\pm 6$ eV$^2$ .
This worrisome feature is common to the observations of several groups;
e.g., the Mainz value is $(-39\pm 34\pm 15)$ eV$^2$. It is possible that
there is an undiscovered systematic error floating around in all the
experiments, and when that gets discovered the current limits might become
less stringent.

{\it Event Pile-up}: At the Bangalore meeting, Wolfgang Stoeffl mentioned
the problem of the Livermore group: `there are {\it too many counts} in
the tritium beta spectrum close to the endpoint'. Such a pile-up of events
is also seen in the Troitsk experiment \cite{BEL}. If the excess count is
due to relic neutrino capture $\nu _{\rm relic}~+~^3_1H ~\rightarrow
{}~^3_2He~+e^-+18.5~~{\rm keV}~,$ the relic density would have to be quite
high $\sim 10^{17}~\nu $'s/cc , which is about $10^{15}$ larger than the
value implied by the big bang cosmology. In any case it may not be
meaningful to give a value for $m(\nu )$ until the shape of the beta
spectrum is fully understood. But if one ignores the problem, and uses
statistical procedures blindly, the Livermore limit \cite{BAN} turns out
to be $m(\bar {\nu }_e)\leq 5~$eV.

No one talks about the 17 keV neutrino any more. It has gone the way of
`cold fusion'. All that hubhub of the early 90's is now happily forgotten.
The origin of the spurious signal in Sulphur-35 decay has been traced
primarily to the thickness of the source, improper appreciation of
scattering effects, etc., \cite{BOW}.

{\bf (B)} NEUTRINOLESS DOUBLE-BETA DECAY: So far no one has seen this type
of decay. Here one checks whether the total energy of the two final
electrons $(E_1+E_2)$ has a unique value. Evidence for the $(\beta \beta
)_{0\nu }$ decay implies lepton number violation and possible existence of
Majorana mass of the electron neutrino. From a negative result of the
search, the quantity that can be constrained is the `effective mass' of
the Majorana $\nu _{e}$ ,
\[ m_{ee}~\equiv\sum _{i}~\eta _{i}^{CP}~m(\nu _i)~|U_{\nu _e
\nu _i}|^2~,\]
where $\eta =\pm 1$ denotes the $CP$ eigenvalue of the $i$ th Majorana
neutrino $\nu _i$ whose mass is $m_i$ , and $U_{ab}$ denotes the element
of the unitary lepton-mixing matrix. Assumptions made are $CP$ invariance
in the leptonic sector and absence of right-handed currents in weak
interactions.

The best limit to date on neutrinoless double-beta decay comes from the
radiochemical experiment of the Heidelberg-Moscow group using enriched
isotope $^{76}$Ge: the 90\% CL lower limit on the halflife is \cite{BELL}
\be T^{0\nu }_{1/2}>3.2\times 10^{24} {\rm years}. \ee
The implied limit on the effective mass of the Majorana neutrino is
\be m_{ee}~\leq ~ 0.9~{\rm eV~~(90\%~CL)}~, \label{DOB}           \ee
which is not inconsistent with that in Eq.(\ref{mass}) because we can have
$\eta ^{CP}_i=+1~or~-1$. This group, which hopes to reach a limit of 0.2
eV within five years, also gave \cite{HEI} the most precise value for the
halflife of $^{76}$Ge for 2-neutrino double-beta decay
\be T^{2\nu }_{1/2}(e^-e^- \bar {\nu_e }\bar {\nu_e})~=~(1.42\pm 0.03\pm
0.13)\times 10^{21}~{\rm years}~.                                    \ee
May be, the decays having 2 (anti)neutrinos are the only type of double
beta-decays that will ever be seen.

{}From data on the isotopes Tellurium-128 and 130, the limit quoted by
Ramanath Cowsik and collaborators \cite{COW} is equally tight, $m_{ee}\leq
1.1-1.3$ eV. This comes from a geochemical experiment using mass
spectrometry.

There is a clever suggestion due to Raghavan \cite{RAG} to bring the limit
on $m_{ee}$ further down to values $<{1\over 4}$ eV. His idea is to search
for the neutrinoless double-beta decay of $^{136}$Xe by exploiting the
low-background environment available in the large water mass at the
Kamiokande facility.

{\bf (C)} MUON NEUTRINO: The long standing 270 keV limit \cite{PDG} has
to be replaced now by the limit \cite{ASS}
\be m(\nu _{\mu })~\leq ~160~{\rm keV}~~(90\%~ CL)~~.    \ee

The experiment measured muon momenta in $\pi ^+$ decays at rest. At
present there is a discrete two-fold ambiguity in quoting the charged pion
mass at the level of keV \cite{JEC}. This gives rise to two solutions in
obtaining $m^2(\nu _{\mu })$ ; solution B (`heavy' pion) leads to a mildly
negative value (at $\sim 1\sigma $) and this solution yields the above
bound; solution A (`light' pion) gives a strongly negative m$^2$ value (at
$\sim 6\sigma $) which is rejected. One hopes that the existing ambiguity
in pion mass will be resolved in favour of the `heavy' pion. Ongoing PSI
experiments on pi-mesic atoms with low-Z, like O or N, will be of interest
in this context.

{\bf (D)} TAU NEUTRINO: Recent data from BEPC in China show that some
numbers in tau-lepton physics have errors smaller than those listed in the
1994 PDG tables \cite{PDG}; e.g., the present world average value
\cite{PAT} of the tauon mass is $M_{\tau }=1777.00\pm 0.26$ MeV and the
tauon meanlife is $\tau_{\tau } =0.2908\pm 0.0015~$ ps. Decays such as
$\tau ^- \rightarrow \nu _{\tau }~+$ (5 or more hadrons) are used to
extract the upper limit on the mass of $\nu _{\tau }$
\be m(\nu _{\tau })~\leq ~29~{\rm MeV}~~(95\%~CL);                \ee
this is a small improvement over the earlier limit of 31 MeV set by the
ARGUS group.

{\bf (E)} NUCLEOSYNTHESIS AND $\nu _{\tau}$ MASS: Arguments based on
nucleosynthesis (NS) in the early universe \cite{KOL} and the observed
elemental abundances, allow us to push down the laboratory upper limit of
the $\nu _{\tau }$ mass by an order of magnitude, $m(\nu _{\tau })<0.2$
MeV. According to BBC (Big Bang Cosmology), primordial synthesis of the
light elements occured when the Universe was about 3-minutes old and had a
temperature of $\sim $ 0.1 MeV.  In that era massive neutrinos ($m(\nu
)\simeq 0.1-1$ MeV) would contribute to the energy density a
nonrelativistic component with the maximum occuring for masses $m(\nu
)=4-6$ MeV. This maximal energy density may be looked upon as due to
adding a few more types of massless neutrinos, $\Delta N_{eff}=4-5$. Since
the observed abundance of premordial helium can be used to restrict the
effective number of relativistic neutrinos, we could rule out a range of
neutrino masses.

This idea was refined \cite{KAW} to include the case of decaying
neutrinos. The $\nu _{\tau}$ should not decay too fast but survive till
the period of NS to contribute to the nonrelativistic energy density. For
$N_{eff}<$ 3.3 (suitable for the helium mass-fraction $Y_p\leq 0.24$) and
neutrino proper lifetime $\tau > 10^3$ s, the excluded mass region is 0.1
- 40 MeV. Combining with the limits coming from the laboratory work, the
astro-particle bound may be quoted as \cite{SMI}
\be m(\nu _{\tau })~\leq ~0.1-0.2~ {\rm MeV~~,~for~}\tau _{\nu _{\tau
}}>10^3~{\rm s}~.                                                \ee
This is a very impressive limit as the mass-ratio is $[m(\nu
_{\tau})/m_{\tau }]\leq 6\times 10^{-5}$, which is only a factor 6 larger
than the ratio for the electron, $[m(\nu _e)/m_e]\leq 10^{-5}$.

For unstable neutrinos with $\tau <100$ s, a gap develops between the
upper limit of the He bound and the lab limit. For $\tau <0.01$ s, the
mass region 3-30 MeV is not excluded at all.

The above limit is valid not only for $\nu _{\tau }$ but also for $\nu
_{\mu }$ . Incidentally, the recent observation \cite{SON94} of a higher
abundance of primordial deuterium $(2-3)\times 10^{-4}$ (expected amount
is $\sim 8\times 10^{-5}$) may revise the calculated primordial $He$
fraction but only very slightly, and hence the tau-neutrino mass limits
are expected to be essentially unchanged.

\newpage

\centerline {\bf 4.~NEUTRINO MAGNETIC MOMENTS}
If the standard model is extended to include right-handed neutrinos, then
a Dirac neutrino with mass $m(\nu )$ has (at one-loop order) the
Lee-Shrock magnetic moment
\be \mu ^{LS}(\nu )= {3\over 8 {\sqrt 2}\pi ^2}~eG_Fm(\nu )~\simeq
{}~0.32\times 10^{-18}~{m(\nu )\over eV}~\mu_B~,                      \ee
where $e>0$ and $\mu _B=e/(2m_e)~\simeq 5.8\times 10^{-15}$ (MeV/Gauss) is
the Bohr magneton. This result is independent of the charged lepton mass
$m_{\ell }$ and possible neutrino-mixing strengths $U_{ij}$.

Following are the currently available 90\% CL upper limits on the neutrino
magnetic moments extracted from the laboratory experiments involving
elastic scattering $\nu e\rightarrow \nu e$ :
\be  \mu (\nu _e)<1.08\times 10^{-9}~\mu _B~,\ee
\be \mu (\nu _{\mu })<0.74\times 10^{-9}~\mu _B~,\ee
\be \mu (\nu _{\tau })<0.54\times 10^{-6}~\mu _B~.\ee
The first two bounds are from the results of the LAMPF group
\cite{KRA}, which has the actual experimental constraint in the form
$\sqrt {\mu^2(\nu _e)+2.1\mu ^2(\nu _{\mu })}~<~1.08\times 10^{-9}~\mu
_B~$. As for the third bound, although $\nu_{\tau }$ has not been
identified so far, its elastic scattering was searched for by
examining events containing forward-scattered single electrons in a
beam dump experiment \cite{COO}.  The source of $\nu_{\tau }$ was
taken as charmed-strange meson decays, $D^+_s\rightarrow \tau ^+
{}~\nu_{\tau}$, etc, where the $D^{\pm }_s $ were produced in high
energy nuclear collisions. The bound on magnetic moment, which is an
order-of-magnitude lower than the earlier one, however depends on
assumed values for the production rate and decay-constant of the
$D^{\pm }_s$ mesons. Improvement in the limit is likely from future
data on $Z$ decays to final states having single photons \cite{GOU}.

Limits extracted by using astrophysics information (energy loss rate of
red giants, Supernova 1987A, ... ) are better, but open to argument due to
their model-dependence.

A recent proposal \cite{SAK} to measure the magnetic moment of $\nu _e$
suggests a novel use of Transition Radiation (TR) detectors. TR is emitted
when a charged particle crosses the interface between two different media.
It should be produced even when a neutrino crosses the interface of two
media provided the neutrino has a nonvanishing magnetic moment. The signal
increases with the Lorentz factor $\gamma $. However, even for the large
solar neutrino flux, with the value $\mu (\nu _e)\sim 10^{-10}~\mu _B$ ,
the estimates of Grimus and Neufeld \cite{SAK} show that the yields of
typical experiments are going to be extremely small ($\sim 10^{-4}$
photons per year); hence the idea may not be practicable.

Segura {\it et~al~} \cite{SEG94} noted that the elastic cross section for
$\bar {\nu }_ee^-\rightarrow \bar {\nu }_ee^-$ calculated in the lowest
order vanishes exactly at one point, namely, when the incident neutrino
energy is $E_{\nu }= m_e/(4\sin ^2{\theta _{W}})~\simeq $ 0.55 MeV and the
final electron is emitted forward (carrying maximal kinetic energy $\simeq
$ 0.38 MeV). Hence detection of events at this `dynamical zero' might
signal scattering due to $\bar {\nu }_e$ magnetic-moment. Another method
to look for neutrino magnetic moment is to study, in suitable regions of
phase space, the inelastic radiative process using reactor antineutrinos
\[ \bar {\nu_e}~e^-\rightarrow \bar {\nu_e}~e^-~\gamma ~. \]
But the cross sections turn out to be discouragingly small ($\sim
10^{-47}~cm^2$ at MeV neutrino energies and for $\mu (\nu )\sim
10^{-10}\mu _{B}$ ); see, Bernabeu {\it et~al~}\cite{SEG94}.

\bigskip

\centerline {\bf 5.~SOLAR NEUTRINO RESULTS}

All the 4 experiments which are currently running show deficit of $\nu_e$
flux from the Sun. For details concerning the solar neutrino experiments,
we refer to the excellent review by Cremonesi \cite{CRE93}. In the
following, Standard Solar Model (SSM) results of Bahcall's group and
Turck-Chieze's group will be denoted by the indicies `$B$' and `$T$'.

Solar neutrinos, in increasing order of their energies, fall into 3 broad
classes: pp,~$^7$Be,~$^8$B. Kamiokande group selects events containing
energetic electrons and hence its signal depends only on neutrinos coming
from the $^8$B decay. In the Cl-Ar experiment, out of the expected total
rate, (71-78)\% is due to neutrinos from $^8$B and (15-17)\% is from the
electron-capture by $^7$Be. In the Ga-Ge experiment, only (8-11)\% is due
to $^8$B decay neutrinos, (25-27)\% from the $^7$Be, and more than half
(54-57)\% is due to the low energy neutrinos from pp fusion (C, N, O
reactions contribute the remaining $\sim $5\%).

{\bf (A)} {$^{37}$Cl EXPERIMENT}: Threshold = 0.814~MeV, \\ Calculated
Rate : $R_{SSM(B)}=~8.0\pm 1.0~SNU,~~~R_{SSM(T)}=~6.4 \pm 1.4 ~SNU.$
\\ Observed Rate \cite{LAN}: $R_{obs}~=~2.55\pm 0.17\pm 0.18~SNU,$
\begin{eqnarray}
{{\rm Observed~Rate}\over {\rm SSM~Rate} }&=&0.318\pm 0.051,~{\rm for~B}
\\&=&0.398\pm 0.096,~{\rm for~T~}.
\end{eqnarray}

The original 1987 claim of anticorrelation with the Sun-spot number at
$5\sigma $ level has now become a $2\sigma $ effect. It is now generally
agreed that there is no significant correlation of neutrino flux with
solar activity.

{\bf (B)} KAMIOKANDE: Minimum electron energy detected = 7.5~MeV \\
Calculated~Flux : $\phi _{SSM(B)}=5.69(1\pm 0.15)\times
10^6~cm^{-2}s^{-1}~,~~~\phi _{SSM(T)}=(4.43\pm 1.1)\times
10^6~cm^{-2}s^{-1}~.$

This is the only ``neutrino-telescope" we have at the moment. Neutrino is
detected by the Cherenkov light emitted by the recoil electron in the
reaction $\nu_ee\rightarrow \nu_ee$ . However the `directionality' is
smeared as the recoil electron suffers multiple scattering in water; the
angular resolution is $\Delta \theta \sim 27^0$ at 10 MeV. [Recall that
the celebrated neutrino burst observation by Kamiokande on 23 Feb 1987 had
11 events with energies in the range 7.5 - 36 MeV; they were due to
(anti)neutrinos from Supernova SN 1987A initiating the nuclear reaction
$\bar \nu_e+p({\rm in~H_2O})\rightarrow e^++n$, and not due to their
scattering on atomic electrons].

Preliminary results of the latest phase KAM-III (with electron energy
cutoff lowered from 7.5 MeV to 7.0 MeV) for 627 days are available; when
these are combined with the earlier 1040-day results of KAM-II, the
combined flux for the 1667-day run is \cite{NAK} $$ {\phi_{{\rm
KAM}~(II+III)}}=(2.9\pm 0.2\pm 0.3)
\times 10^6~cm^{-2}s^{-1}~.~ $$
This flux would correspond to the ratio
\begin{eqnarray}
{\phi_{\rm KAM}\over \phi_{\rm SSM}}~
                    &=&0.51\pm 0.04\pm 0.06~,~{\rm for~B}\\
                    &=&0.66\pm 0.05\pm 0.08~,~{\rm for~T}~.
\end{eqnarray}
$\bullet $ No time variation was noticed in the KAM results during Jan
87-July 93. No correlation with solar activity was evident. No evidence
for day-night effect was observed, [(Day$-$Night)/ SSM flux]=$-$0.08 $\pm
$0.08.
\\ $\bullet $ Observed neutrino energy spectrum shows no evidence for any
distortions; it is as expected in the SSM.
\\ $\bullet $ The next generation experiment `Super-Kamiokande' is being
planned with 50 kilo Tons of water.

{\bf (C)} GALLEX: Threshold = 0.233~MeV, \\
Calculated Rate : $R_{SSM(B)}=~131.5^{+7}_{-6}~$SNU~,~$R_{SSM(T)}= ~
122.5\pm 7~$SNU~,\\with the mean value $R_{SSM}= 128\pm 8$ SNU.

The observed rates for two periods of exposure are \cite{GAL94} \\
GALLEX I : 81$\pm $17$\pm $9 SNU (for 324 days during May 91-May 92)\\
GALLEX II: 78$\pm $13$\pm $5 SNU (for 406 days during Aug 92-Oct 93).
\\The combined rate is
\begin{eqnarray}
\bar R_{{\rm GLX}~(I+II)}&=& 79\pm 10\pm 6~{\rm SNU}\\
                         &=& 79\pm 12~{\rm SNU}.       \end{eqnarray}
This number is based on a total of 136 decays of $^{71}$Ge which are
attributable to solar neutrino interactions during the 2-year period.

$\bullet $ The observed rate is significantly different from the SSM
expectation ($\sim $ 128 SNU). However it coincides with the `minimum
stellar model' rate 78-80 SNU. This value is obtained assuming that pp and
pep fusions are the {\it only} sources of solar neutrinos. Note that in
SSM the rate due to pp and pep neutrinos is actually slightly lower at 74
SNU; this is because 15\% of the terminations go via $^3$He+$^4$He fusions
producing only one $\nu_{pp+pep}$ per termination.
\\ $\bullet $ Exposure time of a solar run is 27 days, and each run is
counted for about 6 months ($\sim $ 11 meanlives of $^{71}$Ge).
\\ $\bullet $ GALLEX II counting will end by 1994. Calibration exposure to
$^{51}$Cr neutrinos (emitted by electron-capture and have energies 0.746
MeV (90.1\%) and 0.426 MeV (9.9\%); planned source will have a strength
$\sim $1.7 megacurie emitting $5\times 10^{11}~\nu ~'s~cm^{-2}~s^{-1}$ at
meter distance initially) will also end by 1994. The entire experiment
itself will be over by December 1996. The ultimate statistical accuracy
aimed for is about 8\% .

{\bf (D)} SAGE: The average rate of the Soviet American Gallium Experiment
as reported recently \cite{GAV} is
\be \bar R_{obs}~=~69\pm 11^{+5}_{-7}~{\rm SNU}~.  \ee
This is about 54\% of the average SSM prediction. Note that the errors are
smaller than the ones given in a recent publication \cite{ABD}.

Together with the GALLEX number the mean rate of the two Gallium
experiments combined becomes
\be                R_{{\rm Ga}}=74\pm 8~ {\rm SNU},            \ee
this is 0.58$\pm $0.07 times the average SSM rate.

{\bf (E)} SOLAR THERMOMETER: Estimated value of Sun's central temperature
is $T_c=15.6\times 10^{6}~K\simeq $1.34 keV. A ``thermometer" to measure
$T_c$ would be the 1.3 keV {\it increase in the average energy} of the
$^7$Be solar neutrino line \cite{BAH}. In the reaction $^7$Be+e
$\rightarrow ^7$Li+$\nu_e$ the peak of the neutrino line is at 861.84 keV.
The asymmetric line profile in the Sun is going to shift the peak value to
862.27 keV, due to absorption of continuum energy electrons. Future
experiments, e.g., BOREX, may be able to provide data with the type of
accuracy needed.

{\bf (F)} COOLER SUN WILL NOT HELP: Reducing the central temperature $T_c$
(also called `the astrophysics solution') will not help to understand the
low rates recorded by Kamiokande and Cl-Ar experiments. The argument is
simple: a cooler Sun leads to a bigger suppression of the energetic
neutrinos (the only kind seen by Kamiokande) while the data require the
other way \cite{BLU}. To see this, let us note the approximate dependences
of the neutrino fluxes on central temperature $T_c$ :
\[\phi _B\sim T_c^{18}~,~~~~\phi _{Be}\sim T_c^{8}~.                 \]
The SSM flux of $\nu _B$ will agree with Kamiokande data provided we
reduce $T_c$ from its canonical value by only about (3$\pm $1)\%. On the
other hand SSM says that the Cl-Ar rate should have about 75\%
contribution from $^8$B neutrinos, about 15\% from $^7$Be neutrinos, etc.
Hence the observed large suppression in the Cl-Ar data can be reproduced
by lowering $T_c$ by a bigger amount, (7$\pm $1)\%. The required
reductions in $T_c$ thus do not overlap and hence a cool Sun is not a
tenable solution - independently of the data from Gallium experiments, a
point first emphasized by Bahcall and Bethe.

Note that it is the relative deficit of the Cl-Ar result with respect to
the Kamiokande result which is crucial, and not the separate deficits with
respect to the SSM.

{\bf (G)} SOLAR NEUTRINO DECAY: This possibility as a way to explain the
solar neutrino deficit is ruled out \cite{ACK}, at more than 98\%
confidence level. In SSM the Gallium rate is fed largely by the low energy
pp-neutrinos while the Kamiokande and Cl-Ar rates by the more energetic
neutrinos. If neutrinos were to decay in transit then it is the lower
energy neutrinos which should disappear first; thus the Ga-rate should
suffer a much stronger reduction than what is observed.

{\bf (H)} NEUTRINO OSCILLATIONS AND MSW EFFECT: \\
{\it Nature within her inmost self divides,\\
To trouble man with having to take sides.}~~~~ - Robert Frost

Solar neutrinos can oscillate into other kinds of neutrinos, $\nu_e
\leftrightarrow \nu _{\mu }~,~\nu _{\tau }~,~ \nu _s~~,$ provided
the relevant mixings and mass differences are nonvanishing (subscript `s'
denotes sterile neutrino having no interaction with a terrestrial
detector). We list the results of analysing the solar neutrino data in
terms of simple 2-flavour oscillations.

Vacuum Oscillation: The long-wavelength oscillation (LWO) solution (also
called ``just-so" solution), assumes the wavelength of oscillations in
vacuum to be about a AU [$\simeq 1.5\times 10^{13}$ cm $\simeq
1/(1.3\times 10^{-18}~$eV)] and the mixing angle to be near maximal. Such
a possibility is not completely ruled out; a typical fit (which gets
allowed at about 10\% CL) for $\nu_e \leftrightarrow \nu _{\mu ,\tau }$
oscillations \cite{KP} is characterized by the parameters
\be \Delta m^2\sim (0.55-1.1)\times 10^{-10}~{\rm eV}^2~~,~~
\sin ^2(2\theta )\sim 0.75-1~.~~                                 \ee

The case of LWO into sterile neutrinos ($\nu _e\leftrightarrow \nu _s $)
cannot be distinguished from the case of ($\nu_e \leftrightarrow \nu _{\mu
,\tau }$) by the Cl and Ga data. However the results of Kamioka-II
spectral data are able to rule out vacuum oscillations into sterile
neutrinos at 98\% CL \cite{KP}. (Oscillations into the sterile kind are
ruled out even when the MSW effect is included.)

The vacuum solution is testable by the seasonal variation (Earth-Sun
distance changing by $\pm 3.5\%$ during the year) in the intensity of the
monochromatic $^7$Be neutrino line. BOREXINO would be ideal for this type
of work. Since the oscillation length ($\ell =4\pi E/\Delta m^2$) is
proportional to neutrino energy $E$ one should see wiggles in the $^7$Be
neutrino flux as the distance varies. Also useful will be similar
observations at SNO (Sudbury Neutrino Observatory in Canada using
ultra-pure D$_2$O) in which the neutrino produces electron by the reaction
$\nu_ed\rightarrow ppe^-$, and the electron energy ($E_e \simeq E_{ \nu
}+m_D-2m_p \simeq E_{\nu }-$ 0.931 MeV) gets determined by the Cherenkov
light.

MSW effect arises from the unequal scattering of the $\nu _e$ and $
\nu _{\mu ,\tau }$ on electrons. As the solar neutrino passes through the
dense layers of Sun's medium, due to coherent scattering on atomic
electrons, there can be resonant or enhanced conversion of $\nu_e$ into
other flavours. The attractive feature is that the solar $\nu_e$ flux can
be significantly suppressed even for insignificant mixings, like the ones
encountered in the quark sector; obviously there is no need for
fine-tuning of the mixing angles.

For the simple case of 2-neutrino mixing, there are two solutions which
are distinguished by the mixing angle. They consist of allowed values of
parameters confined to certain small regions. We quote the central values
of these island regions from the 1993 analysis of Hata and Langacker
\cite{HL1}:
\be {\rm MSW-1~(Small~Mixing):~~}\Delta m^2\sim 6\times
10^{-6}~eV^2~,~~\sin ^2(2\theta )\sim 0.007~;\ee
\be {\rm MSW-2~(Large~Mixing):~~}\Delta m^2\sim 9
\times 10^{-6}~eV^2~,~~\sin ^2(2\theta )\sim 0.6.~\ee
These are indicative of neutrino masses in the milli-eV range, and
correspond to (vacuum) oscillation lengths $\sim {\cal O}(10^2)$ km for
neutrino energies in MeV range. MSW-1 gives a much better fit than MSW-2.

A word about MSW-2 solution. The $^8$B neutrinos can undergo MSW
conversion, in the Earth's core and mantle. This would allow
`reconversion' inside Earth, resulting in an enhanced night-time signal.
For this reason in fitting to the KAM-II day-night data, the parameter
range of MSW-2 becomes further restricted.

The Vacuum or MSW solutions can also be reconciled with the possible
existence of a Majorana neutrino with effective mass $m_{ee}\sim
0.1-1.0$~eV, \cite{PET}.

{\bf (I)} EXPERIMENTAL LIMITS ON $\nu _e$ OSCILLATIONS : What is the
evidence for neutrino oscillations using laboratory neutrino sources ?
New limits on the oscillation parameters of reactor neutrinos have been
provided by the Kurchatov group \cite{VID}. The experiment measured
inverse beta-decays using the antineutrino flux from 3 reactors (of the
Krasnoyarsk group), almost identical power-wise and situated at distances
57.0, 57.6, and 231.4 meters from the detector. The reactors were switched
on or off selectively, in several combinations, during the course of the
experiment.  The final results quoted are
\be \Delta m^2~< 7.5\times 10^{-3}~eV^2~,~{\rm for}~\sin ^22\theta =1~,\ee
\be \sin ^22\theta <0.15~,~{\rm for}~\Delta m^2~>5\times 10^{-2}~eV^2~.\ee
These are slightly better than the earlier Goesgen reactor limits.

A novel proposal to look for neutrino oscillations \cite{SEG2} exploits
the `dynamical zero' in the differential scattering cross section using
reactor neutrinos on electron targets, $\bar \nu _e~e^-\rightarrow \bar
\nu _e~e^-$ . Recall that this zero occurs for incident antineutrino
energy $E_{\bar \nu _e}= m_e/(4~sin^2 \theta _{W})$ at the neutrino
scattering angle $\theta _{CM}=\pi $; it is absent in other reactions,
e.g. $\bar \nu_{\mu }~e\rightarrow \bar \nu_{\mu }~e$ . Oscillations of
$\bar \nu _e\leftrightarrow \bar \nu _{\mu }$ would then lead to
characteristic changes in scattering rate as one scans past the `zero',
and also by varying the detector distance.

{\bf (J)} WHERE ARE THE~ $^7$Be NEUTRINOS ? Recently there have been many
model-independent analyses of the solar neutrino data incorporating the
luminosity constraint \cite{DEG} - \cite{PAR}. The end result is that for
obtaining any kind of fit to the existing data, the $^7$Be neutrino flux
should be severely suppressed, $\phi _{Be}\simeq 0$. This startling
conclusion obviously poses a serious challenge to the SSM; various brands
of SSM have small changes in their input parameters and those can hardly
cause an order-of-magnitude suppression of a major neutrino flux, second
only to $\phi _{pp}$.

The assumptions underlying these analyses are general :
\begin{description}
\item[\rm (a)] The Sun is
powered by the pp-chain and the CNO-cycle.
\item[\rm (b)] The Sun is in a
quasi-steady state having the same value of ${\rm L}_{\odot }$ for at
least the past few million years. This enables us to connect ${\rm
L}_{\odot }$ with the neutrino fluxes $\phi _{i}$.
\item[\rm (c)] Some of the
minor neutrino fluxes are taken into account: the flux from $pep$ fusion
is assumed to be scaled as $2.4 \times 10^{-3} \times \phi _{pp}$,
following SSM; the $^{15}$O and $^{13}$N neutrino fluxes are related to
each other in the SSM ratio 0.85 : 1 ; neutrinos from $^{17}F$ decay and
from $hep$ reaction are ignored.
\item[\rm (d)] Solar neutrinos reaching the earth
are standard $\nu _e$ having no mass, no mixing, no magnetic moment, etc;
they interact with the detector elements according to the Standard Model
of particle physics.
\end{description}

Assumptions (a) and (b) seem to be quite innocuous and almost mandatory,
while (c) is unlikely to play an important role in determining the major
neutrino fluxes; a popular way to renounce the assumption (d) involves
neutrino oscillations.

The luminosity constraint is a linear relation between the luminosity
${\rm L}_{\odot }$ and the neutrino fluxes $\phi _i$ ,
\begin{eqnarray}
K\equiv {{\rm L}_{\odot }\over 4\pi d^2}&=& 8.54 \times 10^{11}~~{\rm MeV
{}~cm^{-2}~s^{-1}}\\
&=& \sum_{i}~({Q\over 2}-<E_i>)~\phi _i~.
\end{eqnarray}
Here $K$ is the usual solar constant measured to $\pm 0.2\%$ , $Q=26.731$
MeV is the total energy released in the process $4p+2e\rightarrow ~ ^4He+2
\nu _e$ , and $<E_i>$ is the average neutrino energy in the reaction
labelled $i$. We absorb the dimensions of the flux by defining
\be f_i\equiv  {\phi_i \over (10^9~~cm^{-2}~s^{-1})}~,           \ee
and obtain the luminosity constraint (inclusive of the N, O neutrino
contributions) as \cite{HL}
\be f_{pp}~+f_{pep}+0.958~ f_{Be}+0.46~f_B +0.955~f_{NO}= 65.7~ .\ee

The neutrino capture rates of interest are given by \cite{DEG}
\begin{eqnarray}
R_{Ga}&=&(79.75 + 2.43\times 10^3~f_B + 6.14~f_{Be} + 7.49~
f_{NO})~{\rm SNU},\\
R_{Cl}&=& (0.247 +1.09 \times 10^3~ f_B + 0.236~ f_{Be}+ 0.396~
f_{NO})~{\rm  SNU}~.                                     \end{eqnarray}
The large factor in front of $f_B$ reflects the large cross section of
$^8$B neutrinos. The flux of $^8$B neutrinos as directly measured by the
Kamiokande experiment is
\be f_B= (2.9\pm 0.36)\times 10^{-3}~.                  \ee

Substituting $f_B$ in the expression for $R_{Cl}$ and ignoring the last
two terms, we get a lower bound which, at 90\% CL, is $R_{Cl}\ge 2.76$
SNU. Since this limit is essentially saturated by the Cl-Ar experimental
result $2.55\pm 0.25$, we immediately see that it is difficult to
accommodate any $^7$Be neutrino contribution (it is about 1.0-1.2 SNU in
SSM). So, where are all the $^7$Be neutrinos gone? Kwong and Rosen
\cite{ROS} had reached this puzzling conclusion using slightly different
arguments.

The recent analyses consist in determining the 4 unknown fluxes $f_{pp},~
f_{Be},~f_B, \break f_{NO}$ from the 4 experimental data $R_{Ga},~R_{Cl},~
f_{Kam},~ L_{\odot } $ ; this is a case with zero degrees of freedom. The
best fit according to chi-square minimum is found to occur for a negative
$^7$Be flux \cite{DEG,HL,PAR}. On the other hand imposing the physical
requirement $f_i \ge 0 $ leads to fits which are quite poor (large
chi-square) and unacceptable; the $Be$ neutrino flux $f_{Be}$ in that case
hovers around zero (less than ${\cal {O}}(10^{-1})$ of the SSM value).

This conclusion about $\phi _{Be}$ suppression is quite robust: even if we
disregard the data from any one of the three solar neutrino experiments,
the resulting fit shows no significant improvement. A simple way to see is
to rewrite the expressions for $R_{Ga}$ and $R_{Cl}$ as follows:
\begin{eqnarray}
f_{Be}+1.18~f_{NO}&=& 0.18~G-0.40~H-14.1~=~-1.9\pm 1.5~,\\
f_{Be}+1.22~f_{NO}&=& 0.16~G-0.40~K-13.0~=~-2.1\pm 1.4~,\\
f_{Be}+1.68~f_{NO}&=& 4.24~H-4.62~K-1.05~=~-3.6\pm 2.0~,
\end{eqnarray}
Here $G,~H~{\rm and}~K$ stand for the data from Gallium, Homestake and
Kamiokande experiments, and the numbers on the extreme right result from
substituting
\be
G\equiv {R_{Ga}\over SNU}=74\pm 8.5,~~~H\equiv {R_{Cl}\over SNU}=2.55
\pm 0.25,~~~K\equiv 10^3f_B=2.9\pm 0.36~.                          \ee
The flux-sums have negative central values (and even with errors they are
much smaller than SSM expectations) for all the three pairings of the
data: $GH,~GK,~HK$. Thus one is led to conclude that at least two of the
three experiments must be wrong \cite{DEG,HL,PAR}; e.g., the systematic
errors could have been grossly underestimated in two experiments.

A comparison of the data with any SSM now becomes almost an irrelevant
exercise: all SSMs depend on the assumptions (a),(b),(d), while (c) may
not be crucial; differences between them arise largely due to the
differences in values adopted for fusion cross sections, opacities, etc,
or in inclusion of some effects such as diffusion. The discrepancy in the
Beryllium neutrino flux is such that any SSM gets ruled out by,
conservatively speaking, more than 3$\sigma $.

On the other hand if we believe in the validity of all data, then we
should reexamine the basic assumptions (a) - (d) of the analyses. A simple
and attractive way out is to modify the assumtion (d) to include the
effects of possible neutrino oscillations in extracting the solar neutrino
fluxes. A model-independent analysis of the data with neutrino
oscillations, including the MSW effect, seems to place only a weak
restriction on the fluxes \cite{HL}.

\bigskip

\centerline  {\bf 6.~ATMOSPHERIC NEUTRINO ANOMALY}
Atmospheric neutrinos were detected first by the KGF group in August 1965.
They are produced in the atmosphere from the decay of secondaries like
pions, kaons, muons, ... which are produced in the interactions of cosmic
rays with target nuclei like N, O in the atmosphere. The cause for present
excitement is that there is a deficit in one kind of these atmospheric
neutrinos, and this `anomaly' may be the long-sought evidence for neutrino
oscillations $\nu _{\mu }\leftrightarrow \nu _{\tau ,e }$ .

The anomaly was first claimed in 1988 by the Kamiokande group in respect
of their `fully-contained' events. They are due to single particles which
are created and absorbed inside the detector. The single particle could be
a $\mu $ or $e$, produced by the appropriate neutrino in the energy range
say 0.3-2 GeV, and propagating `upwards' through the Earth's medium
(direction is fixed by the orientation of the Cherenkov cone). An event is
termed `$\mu $-like' if the associated Cherenkov ring is `sharp', and
`$e$-like' if the ring is `fuzzy'. Bremsstrahlung from electron creates
secondary $e^+e^-$ pairs which also could give Cherenkov radiation in
water. Hence the interactions of $\nu _e$ or $\bar {\nu }_e$ giving $e^-$
or $e^+$ , are associated with diffuse or fuzzy rings.

In the atmosphere, due to decay chains like $\pi ^+\rightarrow \mu ^+
\nu_{\mu }$ and $\mu ^+\rightarrow e^+\bar \nu_{\mu }\nu_e$ , etc., one
expects approximately $(\nu_{\mu }+\bar \nu_{\mu }):(\nu_e+\bar \nu_e)\sim
$ 2:1; but observations show this ratio to be anamolously small,
essentially like 1:1. This finding, termed the {\it atmospheric neutrino
anomaly} (ANA), was originally seen in events at low momenta ($p<$500
MeV/c) of the produced $e$ and $\mu $, and showed no correlation with
zenith angles.

The ratio of the observed $\mu $-like to $e$-like events is compared with
that obtained from Monte Carlo simulations. The following ratio-of-ratios
which ought to be compared with unity, are taken from the talk of K.
Nakamura \cite{BAN}
\begin{eqnarray}
R_{atm} \equiv {(\mu -{\rm like}/e-{\rm like})_{{\rm OBS}}
\over (\mu -{\rm like}/e-{\rm like})_{{\rm MC}}}
&=&0.60^{+0.06}_{-0.05}\pm 0.05~,~({\rm KAM})
\\ &=&0.54\pm 0.05\pm 0.12~,~({\rm IMB}-3)
\\&=&0.69\pm 0.19 \pm 0.09~,~({\rm SOUDAN}-2)
\\&=&0.87\pm 0.16 \pm 0.08~.~({\rm FREJUS})   \end{eqnarray}

{\em Remarks}: ($a$) Recent (preliminary) checks carried out with $e,\mu ,\pi $
beams, produced at the KEK accelerator and sent through a big water tank,
give credence to the Kamioka classification of the Cherenkov rings in the
momentum range 0.2-1 GeV/c \cite{GLA}. ($b$) Although agreement among the
calculated atmospheric-neutrino fluxes is poor, especially at low
energies, there is consistency (to within 5\%) among the calculated {\it
flux-ratios} at all energies. ($c$) KAM data agree with IMB on the $\mu
$-like events, but disagree by 2$\sigma $ on the $e$-like events.  ($d$)
IMB group claims, in their upward-going $\mu $-like events, the ratio of
`stopping' muons ($\leq $ 10 GeV) to `through-going' muons is as expected
{\it without} invoking oscillations. ($e$) The anomaly is supported by the
SOUDAN-2 track-calorimeter which has good resolution and descrimination
for e and $\mu $.

{\sf NEW RESULTS}: Kamiokande group has recently reported \cite{FUK94}
data on the fully-contained (FC) and the partially-contained (PC) types of
events. These are initiated by multi-GeV neutrinos with E(visible) $>$
1.33 GeV and correspond to a mean neutrino energy = 5 - 7 GeV (the earlier
sub-GeV data corresponded to $\sim $ 0.7 GeV). Averaging over the zenith
angles the ratio-of-ratios turns out to be not different from the earlier
sub-GeV data
\be R_{atm}= 0.57^{+0.08}_{-0.07}\pm 0.07~.                       \ee
However a significant feature of the new data is the {\it zenith angle
dependence} (sub-GeV data showed isotropy): $R_{atm}$ steadily increases
from $\sim $ 0.3 to $\sim $ 1 as we go from the `upward' ($\theta =\pi $)
events to the `downward' ($\theta =0$) ones.

Just the observation of a zenith-angle dependence itself is a sufficient
reason to invoke oscillations. Considering the extreme cases, neutrinos
travelling downward may have too short a travel length (20 km length of
atmosphere) and hence do not oscillate; those travelling upwards ($\sim
10^4$ km across the Earth) have a very long flightpath available and
oscillate many times; assuming maximal mixing and averaging over many
oscillations would lead to a 1:1 flavour mixture of the upward movers. The
zenith-angle isotropy in the sub-GeV data could be understood as due to
the large uncertainty with which the incident neutrino direction is
ascertained for a given direction of the scattered lepton. Analysis of all
data (including sub-GeV events) \cite{FUK94} gives the allowed region, for
$\nu_{\mu }\leftrightarrow \nu_{\tau }$ oscillations, around the best-fit
values:
\be
\Delta m^2 \simeq 1.6\times 10^{-2}~eV^2~,~\sin^2 2\theta \simeq 1~, \ee
oscillation length will be ${\cal {O}}(10^3)$ km for neutrinos of a few
GeV. The region for the parameters of $\nu_{\mu }\leftrightarrow \nu_{e
}~$ oscillations is roughly the same, only slightly bigger. However in the
latter case since $\nu_e$ 's are involved, matter effects arising from
their travel through the Earth's medium may have to be included; see, e.g.
FIG.1 of Ref. \cite{DS84}.

\bigskip

\centerline {\bf 7.~CONCLUSIONS}

Very few and tiny improvements have occured in the upper limits on
neutrino masses. In the near future, dramatic improvements in the
lab-based experimental limits are unlikely to occur. In tritium beta decay
Mainz group's study within the ``gap" between the 1S-2S atomic levels
in $(^3{\rm He})^+$ ion, might be interesting.

Two CERN experiments, CHORUS (consisting of 800 kg emulsion) and NOMAD
(with electronic detectors) using the neutrinos from CERN SPS, are already
underway looking for $\nu_{\mu }\leftrightarrow \nu_{\tau }$ oscillations,
with a source-detector distance of 800 meters (sensitive to $\sim $1 eV
neutrino masses). There are proposals to look for oscillations by
`long-baseline' studies - shooting accelerator neutrino beams at detectors
located faraway - Fermilab to Soudan mines (730 km), CERN to Gran Sasso
tunnel (732 km); the former needs the Main Injector installed in the
Tevatron, and the latter needs LHC.

As for the solar neutrinos, the present data (stretched to 90\% CL limits)
can be summarized crudely by the hierarchy
\[0.24\leq \left({{\rm OBS}\over {\rm SSM}}\right)_{{\rm Cl}}
      \leq \left({{\rm OBS}\over {\rm SSM}}\right)_{{\rm KAM}}
      \simeq \left({{\rm OBS}\over {\rm SSM}}\right)_{{\rm Ga}}
      \leq 0.70~;                                                    \]
so it is still the Cl-Ar result which poses the most acute challenge to
the SSM. Model-independent determinations of the solar neutrino fluxes
imply that the $^7$Be neutrino component should be totally absent. An
attractive way out is to assume neutrino oscillations. All data can be
understood in terms of neutrino flavour oscillations, with or without the
need of matter effects. We await the next generation experiments: SNO (by
1995 end), Super-Kamiokande (by 1996 end), Borexino, and ICARUS.

The earlier claim of Kamiokande observing a deficit in the atmospheric
$(\nu_\mu +\bar \nu_\mu )$ flux, has firmed up further. The measured ratio
of mu-like to e-like events is about half the expected value. The zenith
angle dependence of this ratio in the multi-GeV neutrino events is the
novel feature of the 1994 data. The Kamiokande group seems to be again
lucky, and may have discovered neutrino oscillations. The final verdict on
this important issue may have to await the long-baseline studies.

With so many experiments in progress, and many yet to start, neutrino
physics should be as exciting as ever, for a long time to come.

\end{document}